# PCNM: A New Platform for Cellular Networks Measurements and Optimization

Tayeb Lemlouma
Department of Networks and Telecommunications
IUT of Lannion (University of Rennes 1) / IRISA
Lannion, France
tayeb.lemlouma@univ-rennes1.fr

Yoann Lefebvre and Frédéric Cespedes
IUP GMI
LIA / University of Avignon
Avignon, France
{yoann.lefebvre, frederic.cespedes}@univ-avignon.fr

*Abstract*—In this paper, we present PCNM, a new mobile platform for cellular networks measurements. PCNM is based on a set of techniques that tailors theoretical calculations and simulations to the real cellular network environment. It includes: (a) modules that measure different parameters of a base station (BS) such as localization, cells identification, time advance information, reception level and quality, (b) a new protocol that optimizes the task of network measurement by monitoring a set of mobile nodes and finally (c) the ability to extend an existing cellular network by adding new base stations. We evaluate our genetic algorithm used to reduce the nodes mobility and optimize the measurement extraction of $N$ base stations using $k$ mobile sensors ($k \geq 1$). We show how connecting real measurements (using mobile sensors in a collaborative way) to theoretical and prediction methods is of high benefits for cellular networks maintenance, extension and performances evaluation.

## I. Introduction

Mobile cellular networks are faced to two major challenges: tremendous demand of mobile users and the constraints of radio resources which are extremely limited. The majority of cellular technologies are based on a set of techniques that circumvents radio constraints and improves the network capacity, coverage and frequency use. For instance, time division multiple access (TDMA) and code division multiple access (CDMA) technologies aim to better exploit the radio spectrum by allowing multiple subscribers to share the same channel. Each specifies the strategy of bandwidth spectrum allocation during a cellular call in order to allow the maximum number of calls to simultaneously take place in a cell. Serving the maximum number of customers and enlarging the network coverage is of paramount importance for cellular network operators. This must be done with keeping the best quality and grade of service (QoS and GoS) in order to guarantee that the network operates in optimum conditions. Such challenges require methods that start from real needs, reflect a real image of the network state and apply efficiently the required tasks for best performances. A number of methods and simulations have been proposed for efficient cellular networks deployment. Tutschku *et al.* [1] propose a tool that considers mobile subscribers and radio transmission based on a Fast-Ray-Tracing (FRT) method for field strength prediction. BSs positioning usually considers the radio signal properties by the mean of statistical assumptions to estimate the field strength [2] or approximations using wave propagation models [3]. Lieska *et al.* [4] discuss simultaneous optimizations of coverage and capacity of cellular network based on the transmission powers, combination of base stations and the number of channels in BSs. In [5], a combinatorial approach is proposed based on the analysis of three algorithms GR, GA and CAT. Some recent work [6] compares the ability of multiple objective algorithms for the antenna placement problem.

In real conditions, the estimation of the network parameters and the prediction of radio wave propagation are usually faced to the environment constraints and random characteristics, e.g. interferences intra-cell and between BSs in TDMA and CDMA networks. This requires the adaptation of assumed and predicted behaviors. For instance, in [2] the path loss within free space propagation model -that depends on the transmitting power and the field strength- was extended regarding morphographical properties by the mean of Gaussian random variable. This last represents the variation in average received power. In this paper, we propose a new platform and set of basic techniques that tailor theoretical calculations and simulations to a real cellular network. This is of paramount importance to validate used models and make the network deployment more efficient for mobile operators. Here, the objective is to propose a platform of minimal hardware investments and calculations that enables a cellular network operation with a fast setup and analysis periods. Concretely, the main sub-objectives are to enable: (a) the measurement of base station properties such as the coverage, reception quality and level in a real environment (e.g. with obstacles, interferences, traffic density, etc.), (b) the monitoring and the optimal organization of a set of mobile sensor nodes that extract base stations measurements in a given area, (c) the capture of the global image of the network state in a given period, and finally (d) the planning of base stations according to the extracted measurements by sensor nodes and the so called *demand node maps* [7] given as input.

## II. Organization of PCNM Platform

PCNM is composed of two main categories of actors: a set of mobile nodes and a central node (a kind of sink node). The mobile nodes act like sensors for the cellular network and perform measurements regarding the base stations. The role of the central node is to monitor the mobile nodes, apply optimizations to reduce the nodes mobility and collect a global image of the network state. The central node can hold static information regarding the network such as the coordinates of base stations, types of used antenna, etc. This organization has two main differences with traditional sensor networks. First, the communication between the central node and the mobile nodes are based on an existing platform (e.g. the GPRS network of the operator). The used protocol is discussed in the next section. Secondly, no peer-to-peer communication is possible between mobile nodes. For implementation purpose, a mobile node could be enriched by a light embedded system in order to get positional information and facilitate the communication with the central node. To get positional information, nodes can use a Global Positioning System (GPS) receiver; this is the case of our implementation. Alternately, mobile nodes can use other techniques such as those based on signal strength information.

### III. CELLULAR NETWORK MEASUREMENTS AND MAINTENANCE

To reach the objective of network measurements and maintenance we address the measurement of a given base station, the positioning of mobile nodes and their mobility and the required communication between the central node and the set of mobile sensor nodes.

#### A. Base Station Measurements

One of the important measurement parameters in cellular and telecommunication networks is the quality of received signal. It depends to several conditions and is influenced by many radio propagation phenomena such as reflection and diffraction. When we perform real measurements in a cellular network, we can observe that true values may significantly vary from predicated by radio models and approximations. So it is necessary to enrich existing models by real measurements that make the deployment or the extension of cellular networks more efficient.

We define a *measurement point* as a geographical point where the mobile sensor node has to achieve some network measurements such as signal strength, reception quality and bit error rate (BER). A measurement point can be associated to a given BS or simply to a geographical position in order to test the coverage (or another property) of the cellular network. Based on the coordinates of a measurement points and the request of the central node, a sensor node can move to the point's position and perform the required measurements. The central node could query the sensor for making many measurements. For example, in the case where the number of measurements points is bigger than the number of available sensors. Hence, a sensor node has to move from a measurement position to another. This poses the problem of reducing the mobility of the sensor in order to accomplish the measurement task efficiently. From an optimization point of view, this is equivalent to solve the *delivery travelling salesman* problem. Indeed, the salesman problem is to visit a number of cities; and to find the shortest trip in a way every city is visited only once. So optimizing the mobility of the sensor node requires to find an ordered set of measurement positions that compose the shortest path of all the possible paths. To solve this problem, we use in our approach genetic algorithm (abbreviated GA)[8] for two reasons. First, the original problem belongs to the family of non-deterministic polynomial time algorithms (NP-complete), indeed its complexity order is of $O(n!)$. Secondly, GAs provide good approximations and applies to non-continuous problems which is our case. Our coding scheme is based on the coordinates of measurement positions. The considered population is composed of a set of individuals where each individual represents a path that visits all the measurement positions. An individual is represented by an ordered set that gives a possible order of visits for the sensor node. To avoid the obtained optimal solution losing, we adopt a circle crossover approach and we consider a fitness function, based on the path's length, that supports a minimal mobility of the sensor node. The section of simulation and experimental results discusses the evaluation of our algorithm.

#### B. Sensor Nodes Dominance

Here we consider $k$ mobile nodes that measure the characteristics of $N$ base station. The objective remains similar as the one discussed previously but with multiple of mobile nodes. Our approach is to consider only neighboring measurement points for each mobile sensor node. We use the same genetic algorithm, as previous, and apply it to a set of partitions for a given area. Measurement points are then associated to the closest mobile node, so mobile nodes can move in parallel to their associated points and perform the required measurements. In order to partition the set of base stations we use the concept of *Voronoi* diagram: a fundamental construct defined by a discrete set of points [9]. In 2-dimensional Euclidean space, the Voronoi diagram of a set of discrete points (here, the mobile nodes) partitions a given area into a set of convex polygons such that all points inside a polygone (and therefore all measurement points of the cellular network) are closest to only one discrete point (a mobile node). Formally, if $M$ denotes a set of $k$ mobile nodes in a given area, the dominance of a mobile node $m_1$ over a mobile node $m_2$ is defined as the sub-area being at least as close to $m_1$ as to $m_2$ :

$$\text{Dom}(m_1, m_2) = \{x \in R^2 \mid \delta(x, m_1) \leq \delta(x, m_2)\}$$

where $\delta$ is the Euclidean distance between two points. The dominance of $m_1$ over $m_2$ is the closed half plane bounded by the perpendicular bisector of $m_1$ and $m_2$. The visited positions $p_i$ by a node $m$ is given by the following set:

$$\text{DomBase}(m) = \{p_i \in P \mid \forall r \in M, p_i \in \text{Dom}(m, r)\}$$

where P is the set of measurement points. The implementation of the partitioning algorithm guarantees that a measurement position $p$ belongs to only one mobile dominance even if $\delta(p, m_1) = \delta(p, m_2)$.

In Figure 1 is presented an example of node dominances. Filled circles represent the measurement positions of the cellular network; open circles represent the current position of mobile nodes. After the construction of dominances, each mobile node is responsible to extract the base station measurements in its dominance.

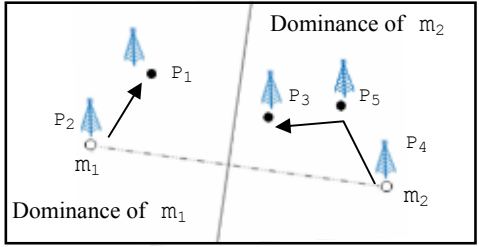

Figure 1. Construction of mobile nodes dominance

*C. Protocol*

The main objective of the high level protocol integrated to PCNM is to monitor mobile sensor nodes and reduce the mobility required for network measurements. Due to the mobility of nodes, the protocol is designed to belong in the stateless family. The light embedded system, introduced in Section II, sends periodically the current position of the mobile sensor node to the central node. This information is used by the central node in order to include, in each messages sent to the sensor, the coordinates of the closest base station for the sensor node. The protocol is mainly based on seven types of messages. Each message represents a query from the central node (respectively, the mobile sensor node) to the sensor node (respectively, the central node) and implies a defined behavior. First five messages identifiers are START, STOP, MOVE-TO, GET-MEASURE and VECTOR. They represent the messages that the central node can send to a sensor node. The last two messages identifiers are READY-TO-SEND and CELL-INFO. They can be sent by a mobile sensor node. Also, the protocol uses a set of control and acknowledgement messages. START and STOP messages are used to begin or end a measurement cycle. MOVE-TO message is used to position a sensor node to given coordinates. GET-MEASURE is used to receive the measurements held by a sensor, the messages format determines if the central nodes is asking for a partial or complete measurements. VECTOR message is defined to give an ordered set of measurement positions that the targeted sensor node must visit. READY-TO-SEND message indicates that the sensor is ready to send measurements to the central node. Finally CELL-INFO message is sent by the mobile sensor node in order to request detailed information about a given cell. Here, the used parameter is the cell identifier that can be measured by the sensor.

IV. IMPLEMENTED MODULES FOR CELLULAR NETWORKS

To measure the cellular network properties and apply previous concepts, we were based on the exploitation of a GSM modem. This is done in order to obtain some experimental results and does not restrict our approach to specific kind of cellular network. Of course, our approach and the simulated modules could be applied to any other generation of cellular network such as UMTS networks. The way in which our implementation for cellular networks was designed is attractive. Indeed, it avoids using many expensive tracers by interrogating directly the network modems (for instance, a GSM modem) and even those embedded on mobile phones. This makes easier to multiply the mobile sensor nodes and exploit them in a collaborative way.

The implementation is composed of a main module who handles the connections between the peripherals (a GSM modem and a GPS receiver) and other graphical modules. These last analyse information data. The communications between the different modules are centralized in a called *communication manager*. The GSM module is exploited either via the traditional serial port or the virtual Bluetooth one if this is supported. The second possible connexion is the one with the GPS receiver via a Bluetooth port.

Positional information are obtained thanks to the GPS receiver, the parsed format is NMEA (National Marine Electronics Association). GSM modem is interrogated via a set of AT commands standardized in the 3rd Generation Partnership Project (3GPP) specifications with additional manufacturer specific commands (SSTK commands). Among the network information that our implementation can measure we cite: cell identifier code, Time Advance, Mobile Country Code (MCC), Mobile Network Code (MNC), Location Aera Code (LAC), reception strength of the current cell and its variation, bit error rate (BER) and its variations, Broadcast Control Channel (BCC), BTS Color Code (BTCC) and Network Colour Code (NCC), see Figure 2.

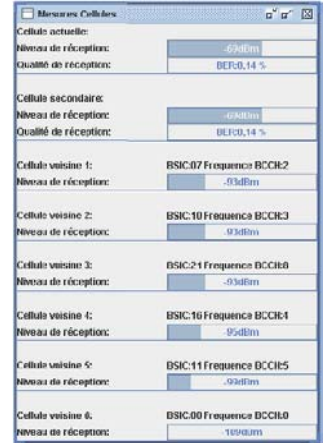

Figure 2. Some measured parameters

For the network extension, i.e. positioning new base stations in positions that are not covered by the exiting network, a similar approach to the one presented in [1] is adopted. Here, our contribution is to correct the demand node map given in input by real measurements done with our platform and applied in some measurement points (Figure 3). Points are selected in the frontier and the intersection of the predicated coverage of the existing network and the input map of demand nodes. Measurements make the decision of maintaining the demand node points or not before starting the planning of BSs.

V. SIMULATION AND EXPERIMENTAL RESULTS

Experimentations done with implemented modules for cellular networks have concerned the measurements of several parameters in different measurement points. Also, the test of the proposed protocol between a central and a mobile node with a GPS receiver. Mobile node measurements have permitted to generate some cartography maps. Figure 3 presents a cartography generated after several measurements in the city of Avignon (France).

Simulated experimentation was done on a test area of 50km*50km with measurement points of random distribution. Simulations have concerned the partitioning of the plan when there is more that one mobile sensor node (i.e. $k > 1$) and the mobility reduction using our GA in the two cases: $k = 1$ and $k > 1$. The speed of mobile sensor nodes (when they move) was set to 30km/h. As it was predicated theoretically, the partitioning generation, using the Voronoi diagram, obeys approximately to the order of $O(n \log n)$ where $n$ is the number of mobile sensor nodes. As we can see from Figure 4, the optimization of mobility works. The time required to perform the measurements is reduced and obtained results are near to the optimum. Also we can notice that the algorithm converge quickly. Indeed, starting from the $130^{th}$ iteration in Figure 4(a) the result approaches the optimum. In Figure 4(b) we can see that with the use of $k > 1$ sensors, the mobility time is significantly reduced. The time of the overall measurements is the maximum of the times required by all the sensors. The sum of times of the 5 sensors is not the same as the time required if we use only one sensor. This is due to two main reasons : (1) the GA who does not give exactly the same result in each execution (2) the union of paths optimized locally in each sensor node dominance does not give the whole path optimized without sensor dominances; this depends to the location of different network partitions and the initial position of mobile sensor nodes.

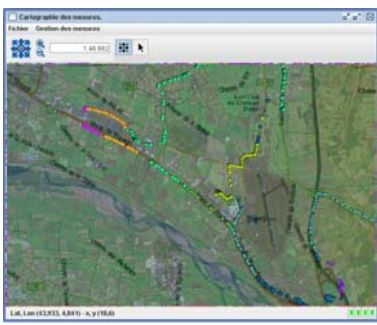

Figure 3. Measurements cartography

## VI. CONCLUSION

Different concepts for cellular networks measurements and maintenance were presented. An optimal and quick algorithm based on GA was proposed and evaluated for performing the network measurements. Also a new protocol was addresses in order to accomplish the capture of a overall network image in a collaborative and efficient way. In addition to the measurement aspects, the proposed platform represents an efficient tool for evaluating, extending or adjusting the configuration of an existing network. PCNM experimentations have not only provided a well knowledge about cellular network characteristics and its variations in different areas but also regarding existing demand nodes that appear because of the population movement, new environment constraints (new obstacles, interferences, etc.) or simply due to the limitations of the prediction models adopted in the original deployment of the existing cellular networks.

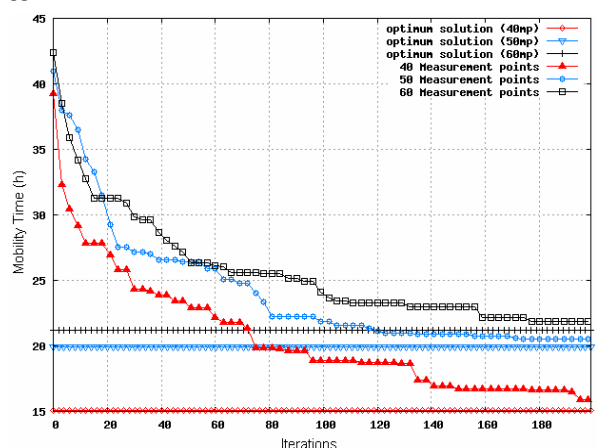

Figure 4(a). Mobility with one mobile sensor node and 40, 50 and 60 measurement points

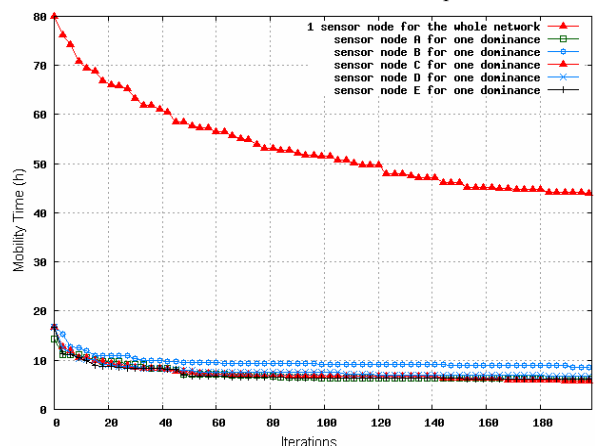

Figure 4(b). Mobility with 1 or 5 mobile sensor nodes and 100 measurement points